\documentclass[aps,prl,twocolumn,showpacs,amsmath,superscriptaddress,floatfix,citeautoscript]{revtex4}

\usepackage[pdftex]{graphicx}
\usepackage{dcolumn}
\usepackage{multirow}
\usepackage{xcolor}
\usepackage{natbib}
\usepackage[colorlinks,bookmarks=false,citecolor=blue,linkcolor=red,urlcolor=blue]{hyperref}
\begin{document}
\title{Hyperbolicity in 2D transition metal ditellurides induced by electronic bands nesting}

\author{Hongwei Wang}
\affiliation{Department of Electrical and Computer Engineering, University of Minnesota, Minneapolis, Minnesota 55455, USA}

\author{Tony Low$^*$}
\affiliation{Department of Electrical and Computer Engineering, University of Minnesota, Minneapolis, Minnesota 55455, USA}

\date{\today }

\begin{abstract}
Naturally occurring hyperbolic plasmonic media is rare, and was only recently observed in the 1T$'$ phase of WTe$_2$. We elucidate on the physical origin of this strong infrared hyperbolic response, and attribute it to band-nested anisotropic interband transitions. Such phenomenon does not occur in general anisotropic materials, at least not in their pristine state. However, band-nested anisotropic interband transitions can in principle be induced via proper electronic band nesting. We illustrate this principle and demonstrate a topological elliptic-to-hyperbolic transition in MoTe$_2$ via strain engineering, which is otherwise non-hyperbolic. 
\end{abstract}

\maketitle
\textit{Introduction.\ }Hyperbolic materials are characterized by anisotropic dielectric constants where the  real parts of the principal components have opposite signs~\cite{poddubny2013, nemilentsau2016}. Such material accommodates modes with extremely large momentum, and  exhibits a very high photonic density of states~\cite{zhang2012}. Hence, such material can support directional polaritons~\cite{low2017} which belong to half-light and half-matter bosonic quasiparticles~\cite{liu2015, kasprzak2006}, enabling a series of potential applications, such as superlens~\cite{pendry2000, jacob2006, smolyaninov2007, fang2005}, invisibility cloaks~\cite{pendry2006}, nanowaveguide~\cite{cortes2012}, and sub-diffractional resonators~\cite{guler2015}. A practical approach creating hyperbolic media is to fabricate metal-dielectric structures~\cite{poddubny2013, noginov2009} in which the components of the effective dielectric tensor can be engineered by tuning the constituent proportion and  geometrical arrangement. Such man-made hyperbolic materials ~\cite{low2017} suffer from high plasmonic loss in metal, complex nanofabrication requirement, optical resolution limited by the feature size, and are not electrically tunable.

Natural hyperbolic material would circumvent the above-mentioned limitations, and enable extreme confinement beyond what is possible with artificial man-made metamaterials. Layered 2D materials offer interesting opportunities on this front~\cite{low2017, basov2016}. A well-known example is hexagonal boron nitride, a naturally occurring hyperbolic material that sustains strongly confined phonon-polariton modes~\cite{caldwell2014, li2015}.  The strong hyperbolic phonon polaritons were also discovered in the van der Waals (vdW) layered semiconductor $\alpha$-MoO$_3$~\cite{ma2018, zheng2018,zheng2019}, with accessible electromagnetic confinements up to two orders smaller than the corresponding photon wavelengths~\cite{zheng2018}. In comparison with the metallic plasmon polaritons, the phonon polaritons possess the virtues of significantly reduced optical losses and much higher quality factors. However active control over phonon polaritons is challenging due to the insensitive  lattice vibrations to external stimulus, which limits relevant applications in switchable nanophotonic devices.

Very recently, hyperbolic plasmonic signatures were experimentally observed in transition metal dichalcogenide (TMD) WTe$_2$~\cite{tang2017, wang2020}, a layered vdW material. The hyperbolic dispersion in the WTe$_2$ thin film was also verified experimentally through far-field infrared absorption measurement~\cite{wang2020}. Natural hyperbolic materials of plasmonic origin are rare, and this being the first experimental demonstration. Its physical origin to-date is also not known. Utilizing density function theory (DFT), we calculate the complex dielectric tensors of  WTe$_2$, and confirm the existence of this near-infrared hyperbolic regime observed in experiments. Through electronic bands and wavefunction analyses, we trace the origin of the emerging hyperbolicity to resonant anisotropic interband transitions via band nesting. We demonstrate this in another telluride material, MoTe$_2$, where its pristine form is non-hyperbolic. Through band nesting engineering via strain, we induce a topological transition in MoTe$_2$ from the elliptic to the hyperbolic regime. Hence, our work established the first electronic structure based approach to the engineering of natural hyperbolic materials and can be generally applied to other anisotropic 2D materials and their heterostructures.
\begin{figure*}[ht]
\flushleft
\includegraphics[width=7.0in]{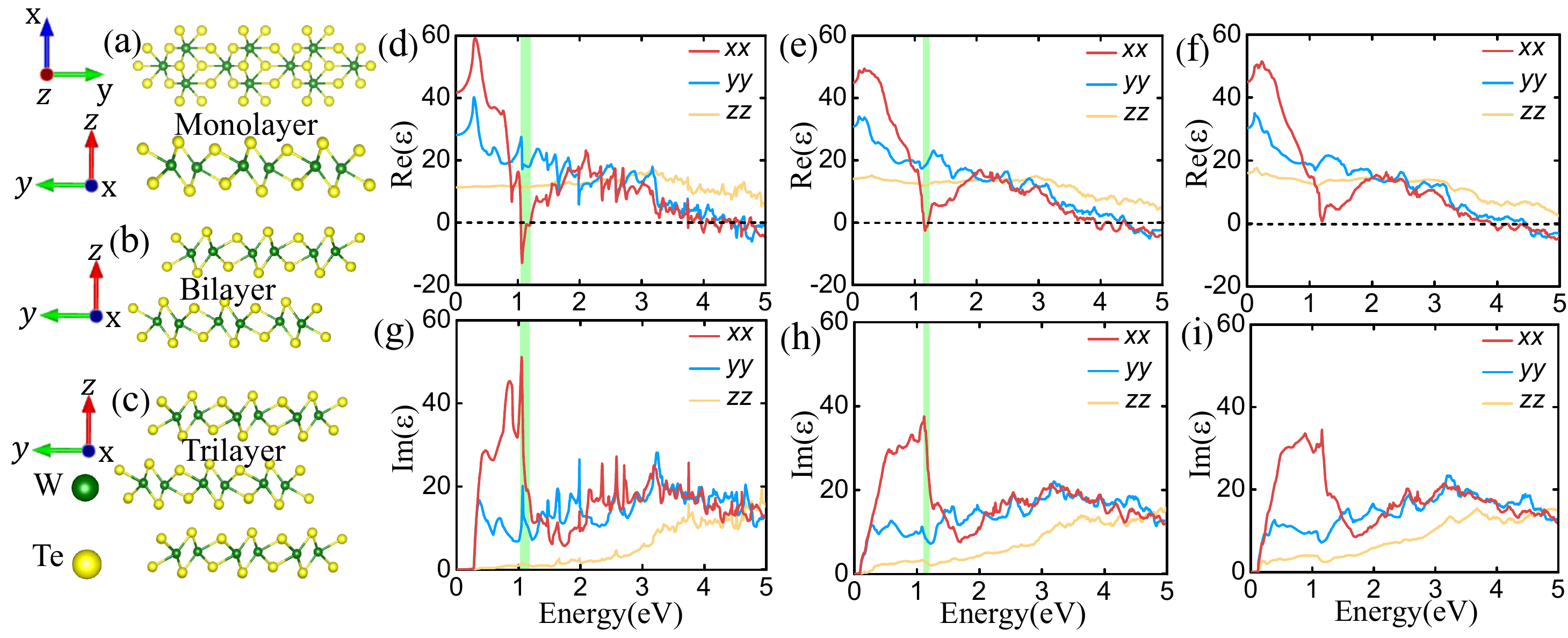}
\caption {(Color online)(a)-(c) Schematic illustration showing the crystal structures of monolayer(a), bilayer(b) and, trilayer(c) WTe$_2$.  The dark green and yellow balls represent the W and Te atoms. (d)-(f) Real parts of the dielectric functions for monolayer(d), bilayer(e) and, trilayer(f) WTe$_2$ materials. (g)-(i) Similar to (d)-(f) but for the imaginary part instead. The spectral region with in-plane hyperbolicity is marked with green background.}
\label{fig:figure1}
\end{figure*}

\textit{Hyperbolic dielectric response.\ }Bulk WTe$_2$ has an orthorhombic crystal structure in its most energetically favorable state, commonly known as the $T_d$ phase~\cite{jiang2016}.
It features monolayer WTe$_2$ structure of the 1T$'$ phase~\cite{fei2017} stacked on top of each other via weak vdW forces. The crystal structures for these layered WTe$_2$ systems are displayed in Figs. 1(a)-1(c), for monolayer, bilayer, and trilayer respectively. The most stable layer stacking order is to rotate the  alternate layers by 180$^o$.
In the simulation, periodic boundary condition is imposed in both in-plane and out-of-plane directions. For the latter, which is denoted as the z-axis, a 20 $\AA$ thickness vacuum layer is inserted to electronically isolate any interactions between supercells.

The calculated real parts of dielectric functions $\varepsilon_{real,j}(\omega)$ for monolayer, bilayer, and trilayer WTe$_2$ are shown in Figs. 1(d)-1(f), and  their respective imaginary parts $\varepsilon_{imag,j}(\omega)$ are displayed in Figs. 1(g)-1(i).  The subscript $j$ denotes the diagonal elements of the dielectric tensor. A material is generally classified as hyperbolic when $\varepsilon_{real,j}\times \varepsilon_{real,i \neq j}<0$. As shown in Fig. 1(d), the dielectric functions of monolayer WTe$_2$ exhibit a strong anisotropic hyperbolic character at $\sim$1.0 eV, such that $\varepsilon_{real,x}(\omega)<0$ and $\varepsilon_{real,y}(\omega),\varepsilon_{real,z}(\omega)>0$. The hyperbolic spectral range resides in the near-infrared and is relatively unchanged, although it narrows, as the layer number increases from monolayer to trilayer as illustrated in Figs. 1(e) and 1(f). The trend continues such that in the bulk limit of WTe$_2$, the material is no longer hyperbolic. This is qualitatively consistent with the experimental observation of hyperbolic response in thin film WTe$_2$ in the infrared spectral range~\cite{wang2020}.

Since the real and imaginary parts of the dielectric function are connected through the Kramers-Kronig relation~\cite{cardona2005}, it suffice to just consider the latter, due to the fact that $\varepsilon_{imag}$ is related to electronic losses, particularly interband transitions. For example, $\varepsilon_{imag}$ with a Lorentzian spectra  with peak absorption at $\omega_0$ would have $\varepsilon_{real}$ which admits a negative permittivity for frequencies larger than $\omega_0$. Indeed, $\varepsilon_{imag,x}$ reveals strong resonant-like features at $\sim$1eV as shown in Figs. 1(g)-1(i). The oscillator strength (or spectral weight) of these resonances would dictates the degree of negative permittivity in $\varepsilon_{real,x}$ as instructed by Kramers-Kronig relation.

As illustrated in Figs. 1(g)-1(i), the peaks at $\sim$1 eV in $\varepsilon_{imag,x}$, which is responsible for the hyperbolicity, becomes weaker and broader as the layer thickness increases. This is consistent with the reduced negative permittivity in $\varepsilon_{real,x}$ as reflected in Figs. 1(d)-1(f). The reducing spectral weight in $\varepsilon_{imag}$ with increasing layer thickness can be traced to the  electronic wavefunction spread between the different layer space ~\cite{li2014}.  As shown in Figs. S1-S3 in Supplementary Materials (SM),   the wavefunctions of the top valence and bottom conduction bands in bilayer WTe$_2$ are more disproportionately localized in one layer and delocalized between the layers, respectively, hence leading to reduced optical transition amplitudes compared to the monolayer counterpart.

\textit{Band nesting effect.\ }To facilitate our discussion of the origin of hyperbolicity in WTe$_2$, we focus on its monolayer and consider its dielectric response between the two bands above and below the Fermi level, denoted as $E_c$ and $E_v$ respectively. As shown in Fig. 2(a), the hyperbolic response is well captured by the optical transitions between $E_c$ and $E_v$ as indicated by the solid lines, where the full response is shown in dashed lines for comparison. As shown in Fig. 2(b),  the intense peak in $\varepsilon_{imag,x}$ at $\sim$1eV can be traced to the interband transitions between $E_c$ and $E_v$. Coincidentally, $E_c$ and $E_v$ display similar dispersions along the $\Gamma-Y$ path, giving rise to a band nesting condition~\cite{carvalho2013} featured by $|\nabla_k(E_c-E_v)|\simeq 0$. A consequence of band nesting is to induce a huge joint density of states (JDOS), since it is defined as ${1\over(2\pi)^2}{\int{dS_k\over|\nabla_k(E_c-E_v)|}}$, where S$_k$ is the constant energy surface defined by $E_c-E_v=$const. Since the spectral weight of $\varepsilon_{imag}$  is proportional to the JDOS, the degree of band nesting in electronic bands of monolayer WTe$_2$ would consequentially control the hyperbolicity. In order to visualize the landscape of band nesting within the Brillouin zone, $E_c-E_v$, $E_c$ and $E_v$ are shown in Fig. 2(c). Besides the $\Gamma$-Y path, band nesting also occurs in other regions of the $\vec{k}$-space as indicated by the iso-energy dashed lines shown in Fig. 2(c). It should be noted that the energy difference between the nested bands is $\sim$1eV, coinciding with the hyperbolic spectral range.

\begin{figure}[ht]
\centering
\includegraphics[width=3.3in]{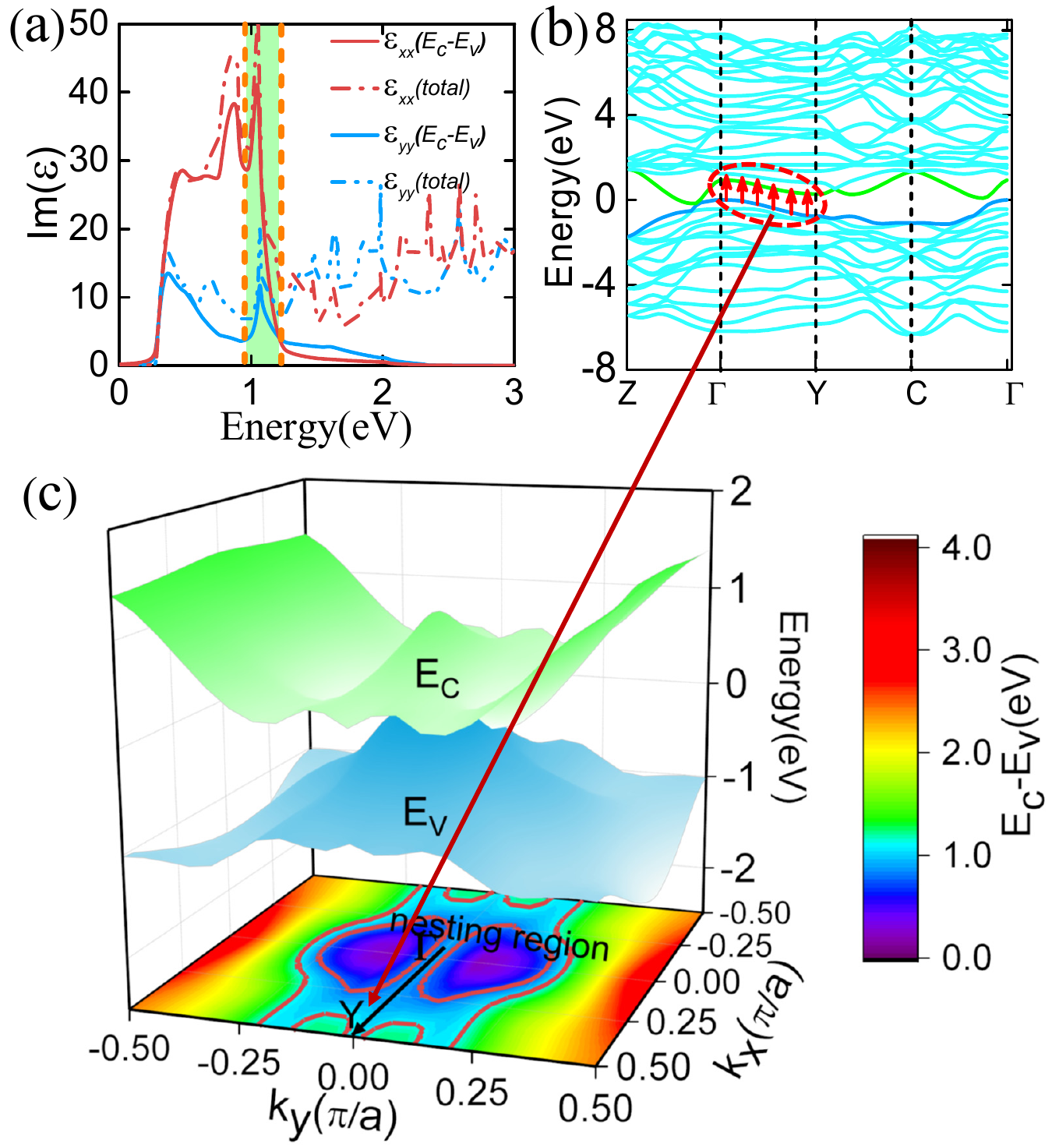}
\caption {(Color online)(a) Imaginary parts of the in-plane dielectric functions for monolayer WTe$_2$.  (b) Two-dimensional band structure along the paths associated with in-plane high symmetry $k$ points for monolayer WTe$_2$.  (c) $E_v$ (blue) and $E_c$ (green) as a function of $k$ mesh. The energy difference between $E_v$ and $E_c$ is shown at the bottom contour projection, with the color bar displayed on right. The lower and upper energy boundaries for the second prominent peak (marked with the green shadow in (a)) in  the imaginary part of dielectric function along $x$ direction are plotted with two contour red lines in (c).}
\label{fig:figure2}
\end{figure}

\textit{Anisotropic opitcal transition.\ }The band nesting phenomenon in monolayer WTe$_2$ is responsible for the `resonant-like' interband transition at $\sim$1eV, and accounts for the appearance of negative $\varepsilon_{real}$ at energy larger than 1eV. However, the anisotropic character of in-plane dielectric tensors is also a pre-requisite to attain hyperbolicity. In fact, the strength of imaginary part of dielectric function is a product of JDOS and the electric dipole transition moment matrix. The electric dipole moment matrix is expressed as $\langle\psi_v|\hat{r}|\psi_c\rangle$, where $\psi_v$ and $\psi_c$ are wavefunctions of $E_v$ and $E_c$, and $\hat{r}$ is the position operator. The symmetry and space orientation of $\psi_v$ and $\psi_c$ would impose constraints on the optical transition selection rules~\cite{yuan2015, hotta2018}, and is responsible for inducing anisotropic dielectric tensors. The $E_v$ band is mainly composed of $p_{z'}$ orbital on Te atoms and the $E_c$ band is formed by $d_{z'^2}$, $d_{x'^2-y'^2}$, and $d_{x'y'}$ orbitals on W atoms (see in Figs. S4 and S5 in SM.). Here the orbitals are defined in the coordinate frame $x'y'z'$ by crystallographic space group notation. The relationships between crystallographic coordinate frame $x'y'z'$ and previous defined conventional coordinate frame $xyz$ are $x\rightarrow z'$, $y\rightarrow y'$, and $z\rightarrow x'$. Due to the symmetry of the wavefunctions, the transition matrix elements $\langle p_{z'}|\hat{r}|d_{x'^2-y'^2}\rangle$ and $\langle p_{z'}|\hat{r}|d_{x'y'}\rangle$ are zero, and only W-$d_{z'^2}$ and Te-$p_{z'}$ orbital projected bands produce finite electric dipole transitions (see Fig. 3). The wavefunctions of $E_c$ and $E_v$ states in the band nesting region are also displayed at the bottom of Fig. 3, which are mainly composed of  Te-$p_{z'}$-like and W-$d_{z'^2}$-like orbitals, in agreement with the maximum localized Wannier functions analysis.
 
 We proceed to discuss the optical selection rules for electric dipole transition. The crystal structure has a mirror plane symmetry define by the $z'$ = 0 plane, and the $E_c$ and $E_v$ wavefunctions in the $\Gamma-Y$ $k$-path exhibit odd and even parities with respect to this mirror plane. As a result, the electric dipole transition is symmetry allowed in $z'$ direction, while the transition in $y'$ direction is forbidden under the mirror plane  restriction. Hence, this selection rule ensures that the band-nested optical transitions occur only for z$'$(or x) polarized excitations. However, this symmetry constraint will be somewhat relaxed in the band-nesting region of k-space off the high symmetry $\Gamma-Y$ path.
Despite this, both Te-$p_{z'}$-like and W-$d_{z'^2}$-like content in the $E_c$ and $E_v$ wavefunction respectively, exhibit strong space orientation in $z'$ direction. Hence, this also attributes to the larger amplitude of transition element $\langle p_{z'}|z'|d_{z'^2}\rangle$ than $\langle p_{z'}|y'|d_{z'^2}\rangle$. These symmetry constraints in conjunction with the orbital type underlie the strong in-plane anisotropic dielectric tensors. Furthermore, we also verified our proposed mechanism in crystal-field coordinate frame which has been widely used to investigate optical transitions in transition metal coordination complexes (see SM text).
\begin{figure}[ht]
\centering
\includegraphics[width=3.5in]{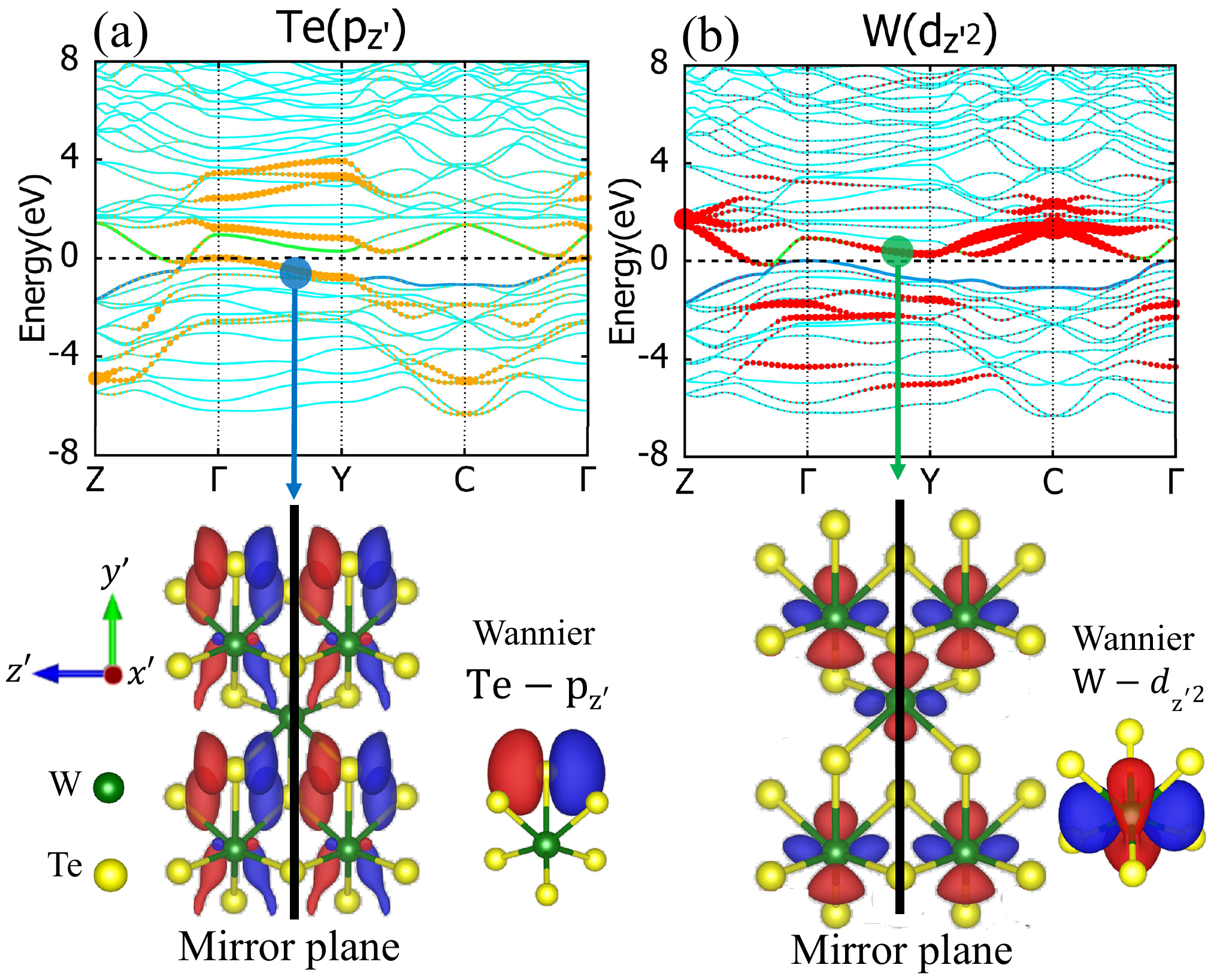}
\caption {(Color online)(a),(b) The projected band structures of monolayer WTe$_2$ with $p_{z'}$ orbital (a) on the Te atom and $d_{z'^2}$ orbital (b) on the W atom. The wavefunctions of $E_v$ and $E_c$ for a momentum along the high symmetry $\Gamma-Y$ line, corresponding to the band nesting region are plotted on the bottom. The Te-$p_{z'}$ and W-$d_{z'^2}$ Wannier orbitals are displayed for comparison.
\\}
\label{fig:figure3}
\end{figure}
\begin{figure}[htb]
\centering
\includegraphics[width=3.3in]{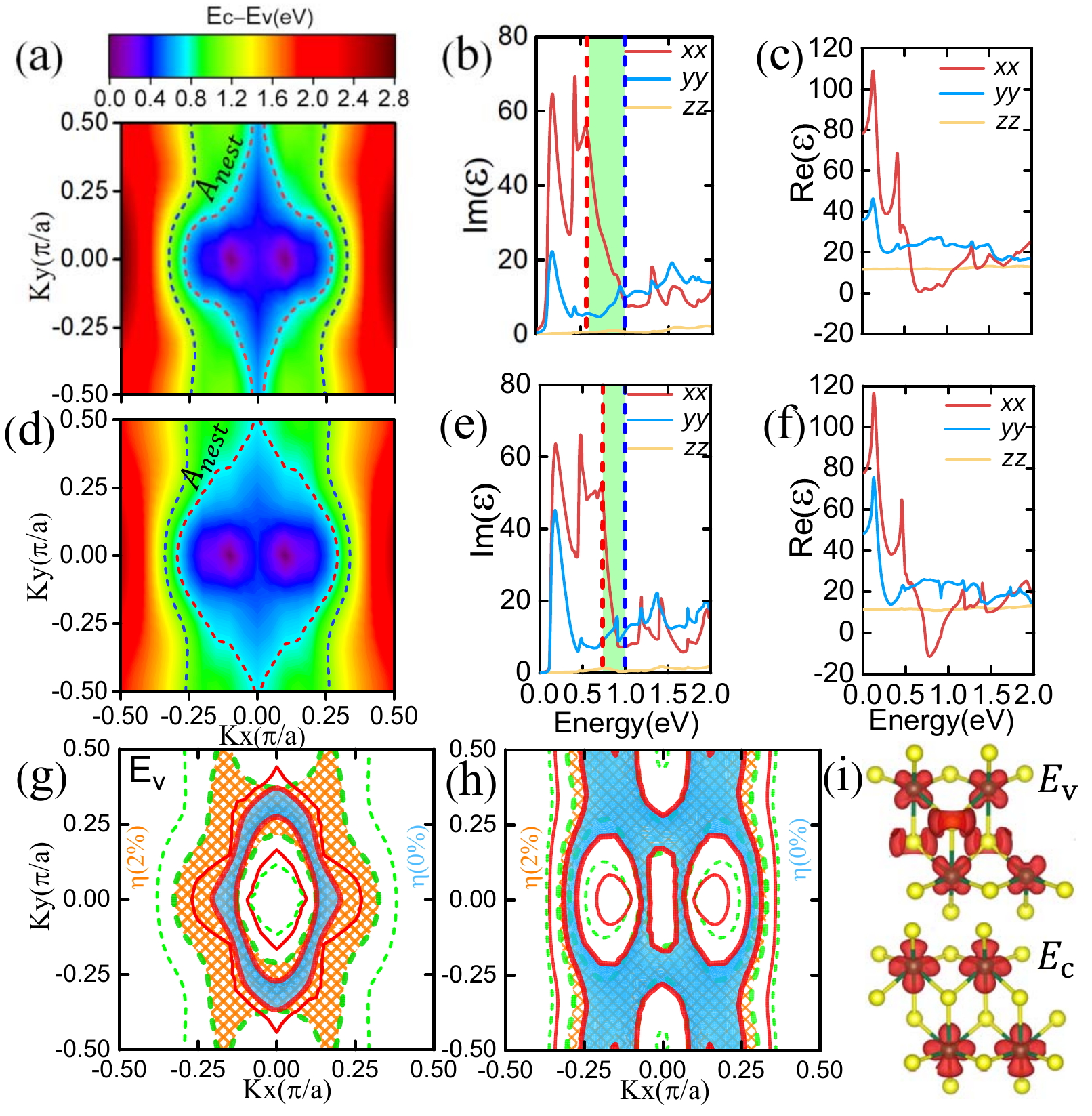}
\caption {(Color online)(a)The contour plot of the energy difference between $E_v$ and $E_c$ for unstrained monolayer MoTe$_2$. (b), (c) Real (b) and imaginary (c) parts of the dielectric functions for unstrained monolayer MoTe$_2$. (d)-(f) Similar to (a)-(c) but for 2\% strained monolayer MoTe$_2$. The two isovalue lines in each contour plot represent the upper and lower energy limits of the shaded green region in the imaginary parts of dielectric functions.  (g), (h) $E_v$ (g) and $E_c$ (h) corresponding to the band nesting region for unstained (blue color) and strained (orange color) monolayer MoTe$_2$. The contour lines in $E_v$ and $E_c$ energy  profiles range from -0.6 to 0.0 eV and 0.0 to 0.6 eV, respectively, with an energy interval  of 0.2 eV. (i) The partial charge densities for the unstained monolayer MoTe$_2$ contributed from electronic states in the band nesting region. The isosurface value is 0.006$e/$$\AA$$^3$.}
\label{fig:figure4}
\end{figure}

\textit{Band nesting modulation.\ }The proposed principle of band nesting induced hyperbolic dielectric response should be generally applicable to anisotropic materials. To demonstrate this, we consider the monolayer MoTe$_2$ system, a sister compound of WTe$_2$, which also possesses stable 1T$'$ phase~\cite{naylor2016}. Figs. 4(a)-4(c) display the band nesting contour for $E_c-E_v$ across the Brillouin zone and the corresponding $\varepsilon_{real}$ and $\varepsilon_{imag}$  for monolayer MoTe$_2$ respectively. Unlike WTe$_2$, monolayer MoTe$_2$ is not hyperbolic in its pristine state across the energy range as shown in green background. Although $\varepsilon_{imag,x}$ also exhibits similar resonant-like features as shown in Fig. 4(b), it is somewhat broadened in energy, as indicated by the green highlighted background. Its lower and upper energy limits correspond to the red and blue contour lines in Fig. 4(a). The center of the second resonant feature  in $\varepsilon_{imag,x}$ of monolayer MoTe$_2$ is localized around 0.6 eV, however the band nesting  corresponding to this energy region is poor, which is manifested by relatively large k-space area bounded by the red and blue contour lines in Fig. 4(a), which herein denoted as A$_{nest}$. Conversely, A$_{nest}$ should be zero for perfectly nested bands across the whole Brillouin zone.

The electronic structure of MoTe$_2$ near the Fermi level is determined by Mo-4$d$ and Te-5$p$ orbitals, and their hybridization is also responsible for the Mo-Te chemical bond. Hence, the electronic structure is tunable through strain engineering. The band nesting contour, $\varepsilon_{real}$, and $\varepsilon_{imag}$ for monolayer MoTe$_2$ under 2\% tensile strain are shown in Figs. 4(d)-4(f). The band nesting profile and dielectric functions of the strained monolayer MoTe$_2$ display evident alterations in comparison with the unstrained one. The strained monolayer MoTe$_2$ exhibits hyperbolic response at 0.8 eV, see Fig. 4(f).  Evidently, the band nesting region around the energy range where the hyperbolicity occurs is appreciably enlarged, while that of A$_{nest}$ is reduced, see Fig. 4(d), resulting in a higher quality peak  in $\varepsilon_{imag,x}$ as shown in Fig. 4(e).

Finally, we examine  how the mechanical strain modulates the band dispersions of $E_v$ and $E_c$ which in turn enhances the degree of band nesting. As indicated in Figs. 4(a) and 4(d), the band nesting does not occur along common high symmetry lines as in the case of WTe$_2$. Hence, we depict the band dispersion of $E_v$ and $E_c$ for the unstrained case across the momentum space of interest as shown in Figs. 4(g) and 4(h) respectively. In particular, we highlighted in blue the region of $k$-space corresponding to the energy window of interest responsible for the resonant-like feature in $\varepsilon_{imag,x}$ and the emergent hyperbolicity. In comparison with $E_c$, $E_v$ band is more dispersive, indicative by the smaller area of the blue shaded region. 

We found that the application of strain has a prominent effect on the dispersion of $E_v$. As shown in Fig. 4(g), the application of 2\% tensile strain appreciably reduces the dispersion of $E_v$, hence increasing the $k$-space area corresponding to this energy window, as indicated by the shaded orange region. The increased in $k$-space area, in conjunction with the relatively non-dispersive $E_v$ and $E_c$ bands, results in band nested interband transitions. The partial charge densities extracted from the electronic states in the above considered energy ranges are shown in Fig. 4(i), which display a mixing of $p$ orbital components on Te atoms and $d$ orbital components on Mo atoms for $E_v$ but only $d$ orbital components on Mo atoms for $E_c$. The band dispersion is determined by the overlap and mixing of atomic orbitals on neighboring crystal sites~\cite{zeier2016}. In monolayer MoTe$_2$,  the $p-d$ orbital mixed component displays a larger overlap between neighboring crystal sites than the one including $d$ orbital only as shown in Fig. 4(i). Therefore, the electronic state of $E_v$ should exhibit a stronger dispersion than that of $E_c$ in our specified energy region. The application of tensile strain reduces the orbital overlap, hence resulting in weaker band dispersion for $E_v$. On the contrary, the interplay of on-site $d$ orbital components related to $E_c$ is less affected by strain.
 
\textit{Summary.\ }In summary, we elucidate the physical origin of the hyperbolicity in WTe$_2$, attributing it to the band nesting effect in conjunction with highly anisotropic optical transition dipole. Guided with this understanding, we applied electronic band nesting engineering to another TMD, MoTe$_2$, and successfully demonstrate elliptic to hyperbolic transition through strain engineering. We envision that natural 2D hyperbolic materials can be made widely available through electronic band engineering via the band nesting approach as outlined in this work.

\acknowledgments
\textit{Acknowledgment.} HW and TL acknowledge funding support from NSF/DMREF under Grant Agreement No. 1921629.$^*$ To whom correspondence should be addressed: tlow@umn.edu.

\bibliography{WTe2}

\end{document}